\documentclass[
twocolumn,
superscriptaddress,
amsmath,
amssymb,
aps,
pra,
]{revtex4} 

\usepackage{graphicx}
\usepackage{dcolumn}
\usepackage{bm}
\usepackage{verbatim}
\usepackage[caption=false]{subfig}
\usepackage{CJK}
\usepackage{color}
\usepackage{enumerate}

\definecolor{myColor1}{rgb}{1,0,0}
\definecolor{myColor2}{rgb}{0,0,1}

\begin{document}
\begin{CJK*}{UTF8}{}
\preprint{APS/123-QED}

\title{Emulation of the dynamics of bound electron exposed to strong oscillatory laser field with Bose-Einstein Condensates}

\author{Ziheng Ma}%
\affiliation{Department of Physics and Institute of Theoretical Physics, University of Science and Technology Beijing, Beijing 100083, China} 
\author{Jia Li}%
\affiliation{Department of Physics and Institute of Theoretical Physics, University of Science and Technology Beijing, Beijing 100083, China}
\author{Rui Jin}%
\affiliation{Institute of Applied Physics and Computational Mathematics, Beijing 100094, China}
\author{Yajiang Hao}%
\email{haoyj@ustb.edu.cn}
\affiliation{Department of Physics and Institute of Theoretical Physics, University of Science and Technology Beijing, Beijing 100083, China}
\date{\today}

\begin{abstract}
This paper employs a Bose-Einstein condensates to simulate the dynamical response of bound electrons in a strongly oscillating pulsed laser field. We investigate the excitation dynamics of Bose-Einstein condensates with repulsive interaction confined in a potential well with finite depth and width driven by a strong oscillatory pulse field. By numerically solving the Gross-Pitaevskii equation with Crank-Nicolson method and split operator method, we obtain the time-dependent wavefunction and therefore the evolution of density distribution in real space and that in momentum space, and the occupation distribution in energy space. It is shown that cold atoms with weak interaction oscillate as a whole body in a finite space when the amplitude of pulse drive is not strong enough. During the evolution atoms occupy the bound states with larger probability. Increasing the driving strength or atomic interactions promotes the excitation of atoms into continuum states and their diffusion out of the potential well, leading to complex structures or even interference-like patterns in the momentum distribution. The number of cycles in the pulse envelope plays a crucial role in the dynamical behavior: High-frequency driving can suppress diffusion and maintain localization. Furthermore, repulsive atomic interactions can enhance high-harmonic generation yields by several orders of magnitude. This study offers a new perspective for quantum simulations of ultrafast dynamics in strong fields and reveals the regulatory role of interactions in condensates on non-equilibrium dynamical processes.
\end{abstract}


\maketitle
\end{CJK*}

\section{Introduction}

With the rapid development of the experimental technique of the cold atom, it has been regarded as a popular platform to simulate interesting interacting quantum systems \cite{NatPhys_8_267, RevModPhys_89_011004}. Not only can the atomic interaction be tuned with the Feschbach resonance and confined induced resonance techniques \cite{FR, CIR}, but also the dimension and geometry of quantum gasses can be controlled \cite{Paredes, Toshiya, Ketterle, Single1D}. Besides the ground state quantum phase \cite{science_329_547, Nature_415_39, RevModPhys_83_1523, NatPhys_8_267}, the quantum dynamics of interacting many body systems are also paid more and more attentions \cite{Bloch2008, RevModPhys_83_863, RevModPhys_85_1191, nature_607_667}, such as the Floquet dynamics in periodical potential \cite{EckardtRMP}, the dynamical localization \cite{DynLoc, DynLoc2}, the dynamical quantum phase transition \cite{DynPhaTran}, modulated dynamics in harmonic trap \cite{EQuinn}, the ultrafast many-body dynamics \cite{Nat_commun_7_13449}, and quantum chemistry simulation \cite{Nature_574_215, PRA_103_043318}.

With the great achievement in the research on cold atoms near equilibrium, the ultracold atom also offer us a nearly ideal context for investigating the quantum systems in the extreme nonequilibrium such as the quantum emulation of ultrafast atom-light interactions---simulating the response of electrons in atoms and molecules to ultrafast strong fields \cite{Ann.Phys.(Berlin)_2017_529_1700008,Krause1992}. In the last decades years, based on the significant theoretical and experimental advances in ultrafast lasers, the detecting and controlling techniques on the microscopic spatial and temporal scale achieve tremendous progress. However, the understanding of nonequilibrium strong-field physics remains to be further in depth. For example, the precise mechanism and timing of tunneling ionization and the effects of interactions are still open questions\cite{Nature_485_343}. Both the numerical precision in theory and the physical limitation in experiment hinder the clarification of these challenges. In particular, the dynamical processes of electrons in atoms and molecules are ultrafast and difficult to detect or control. The tremendous success in quantum emulation of crystal material with neutral atoms in optical lattices prompted the quantum simulation of attosecond science with ultracold atoms \cite{PRA_95_011403,PRA_81_063612}. In a cold atom system, ultrafast processes in strong-field physics such as tunnel ionization evolving on the millisecond scale become observable \cite{PRX_5_031016(2015),NATURE_COMMUNICATIONS_9_2065(2018),PRX_QUANTUM_5_010328(2024),nature_photonics_12_266}.

In the present paper, we investigate the excitation dynamics of Bose-Einstein Condensates (BECs) initially trapped by a Gaussian potential and then driven by a pulse field to mimic the dynamics of the electron in bound state exposed to a strong oscillatory laser field. It is expected that the current quantum emulation of ultrafast processes is complementary to that of conventional ultrafast experiments. The cold atom system to satisfy Bose-Einstein statistics will be investigated, which is very different from electrons to satisfy Fermi-Dirac statistics. Furthermore, unlike the Coulomb interaction between electrons, the atomic interaction is a contact interaction and has exquisite controllability. The ground state and dynamics of weakly interacting BECs are governed by the Gross-Pitaevskii equation (GPE) \cite{Dalfovo1996,PRL_80_3899,PRL_77_2360}, in which the nonlinear terms stem from the $s$-wave scattering between atoms \cite{AlKhawaja2007}. The GPE successfully simulated collective excitation behaviors such as breathing modes and scissors modes in condensed matter systems \cite{Dion2003}. In ultrafast processes, high-harmonic generation (HHG) is an important interacting mechanism, which describes the ionization, acceleration, and recombination of electrons driven by strong laser fields, i.e., the three-step model \cite{PRX_QUANTUM_5_010328(2024),Krausz2009,PRA_49_2117}. It was originally developed for atomic gases and has now been validated across diverse quantum systems including many-electron ensembles, interacting solids, and topological materials \cite{Santra2006, deVega2020, Teubner2009, Faria2001, Faria2002, Yu2024, Amini2019}. So we will investigate the HHG yield in excited BECs and its dependence on the atomic interaction.

We will study the effect of atomic interaction, the strength and frequency of pulse drive on the excitation dynamics of BECs by numerically solving the GPE. The paper is organized as follows. In section II, we present the model and the split operator (SPO) method. In Section III, we present the evolution of density distributions in both real space and momentum space under various driving field and interaction strengths $g$. Section IV illustrates the occupation of continuum states and bound-state energy levels. The HHG yields are presented in Section V, followed by a brief summary in Section VI.

\section{Model and Numerical Methods}
\label{sec:model_method}

We consider a BEC composed of $N$ atoms, each of mass $m$, initially confined by a static external potential $V_{\text{ext}}(x)$ and subsequently excited by an oscillatory pulse drive $V_{\text{drive}}(x,t)$. The system's dynamics is governed by the time-dependent GPE
\begin{equation*}
\begin{aligned}
    i\hbar\frac{\partial}{\partial t}\psi(x, t)=\left(-\frac{\hbar^2}{2 m}\nabla^2+V(x, t)+g|\psi(x, t)|^2\right)\psi(x, t),
\end{aligned}
\end{equation*}
where $V(x,t)=V_{\text{ext}}(x)+V_{\text{drive}}(x,t)$. $g$ is an effective one dimensional interaction strength dependent on the $s$-wave scattering length $a_s$, the atom number in the BEC, the atom mass, and the confinement geometry.

The initial confining potential is a Gaussian well
\begin{equation*}
\label{eq:Gaussian_potential}
V_{\text{ext}}(x)=-V_0\exp\left(-\frac{x^2}{2 r_0^2}\right),
\end{equation*}
where $V_0$ and $r_0$ denote the depth and width of the potential well, respectively. Under the dipole approximation, the driving potential is provided by a strong oscillatory pulse laser and takes the form of
\begin{equation*}
\label{eq:drive_potential}
V_{\text{drive}}(x, t)=x\cdot F\sin^2\left(\frac{\omega t}{2 n_c}\right)\sin(\omega t),
\end{equation*}
which has a period of $T=2\pi n_{c}/\omega$. Here, $F$ represents the  amplitude of the driving field and $T$ corresponds to both the driving period and the total pulse time. The parameter $n_c$ denotes the number of cycles in the pulse field carrier envelope, and $\omega$ is the carrier frequency. To isolate the effect of $n_c$ while maintaining a constant total pulse time $T$, we set $\omega = n_c/4$. This allows us to systematically investigate the influence of the oscillation cycle number $n_c$ and the field amplitude $F$ on the dynamics of BEC, including density distributions, occupations of energy level, and HHG yield. For numerical simplicity, we adopt natural units ($\hbar=m=1$) in all subsequent calculations, reducing the GPE to
\begin{equation}
\label{eq:GPE_natural}
i\frac{\partial}{\partial t}\psi(x, t)=\left(-\frac{1}{2}\nabla^2+V(x, t)+g|\psi|^2\right)\psi(x, t).
\end{equation}
In the following calculation, the depth and width of the Gaussian potential are set to $V_0=10$ and $r_0=2$ in natural units.

\subsection{Ground State Preparation via Imaginary Time Evolution}
\label{subsec:imaginary_time}

The initial ground state of the BEC in the Gaussian potential is obtained by GPE under the imaginary time evolution $\tau=it$. The imaginary time evolution is governed by the equation
\begin{equation}
\label{eq:GPE_imaginary}
\frac{\partial}{\partial\tau}\psi(x,\tau)=\left[\frac{1}{2}\frac{\partial^{2}}{\partial x^{2}}-V(x)-g|\psi(x,\tau)|^{2}\right]\psi(x,\tau).
\end{equation}

We discretize Eq. (\ref{eq:GPE_imaginary}) using the Crank-Nicholson scheme with imaginary time step $\Delta\tau=\Delta$ and spatial step $\Delta x=h$. Applying central differences yields
\begin{equation*}
\label{eq:discretized_imaginary}
\begin{aligned}
\frac{\psi_{j}^{n+1}-\psi_{j}^{n}}{\Delta} = &
\frac{1}{4 h^{2}}\left(\psi_{j+1}^{n+1}-2\psi_{j}^{n+1}+\psi_{j-1}^{n+1}\right) \\
+ & \frac{1}{4 h^{2}}\left(\psi_{j+1}^{n}-2\psi_{j}^{n}+\psi_{j-1}^{n}\right) \\
- & \frac{V_{j}}{2}\left(\psi_{j}^{n+1}+\psi_{j}^{n}\right) - \frac{g}{2}\left|\psi_{j}^{n}\right|^{2}\left(\psi_{j}^{n+1}+\psi_{j}^{n}\right) \\
+ & O\left( \Delta+h^2 \right),
\end{aligned}
\end{equation*}
where $\psi_j^n = \psi(x_j, \tau_n)$. Rearranging terms leads to a tridiagonal system

\begin{equation*}
\begin{aligned}
\label{eq:tridiagonal_system}
-\alpha\psi_{j+1}^{n+1} + (1+\beta_j)\psi_j^{n+1} - \alpha\psi_{j-1}^{n+1} \\
= \alpha\psi_{j+1}^n + (1-\beta_j)\psi_j^n + \alpha\psi_{j-1}^n,
\end{aligned}
\end{equation*}
where $\alpha = \Delta/(4h^2)$ and $\beta_j = \Delta/(2h^2) + \Delta V_j/2 + \Delta g|\psi_j^n|^2/2$. This system can be solved efficiently at each iteration. Propagation to $\tau \to \infty$ yields the ground state wave function.

\subsection{Split Operator Method for Real-Time Evolution}
\label{subsec:split_operator}

We numerically solve Eq. (\ref{eq:GPE_natural}) using the SPO method \cite{Miller2011}. The time evolution operator is approximated by splitting the Hamiltonian $\hat{H} = \hat{T} + \hat{V} + g|\psi|^2$ into kinetic term ($\hat{T}$) and potential term ($\hat{V}_{\text{total}} = \hat{V} + g|\psi|^2$),
and we have
\begin{equation*}
\label{eq:time_evolution_operator}
|\psi(t)\rangle = \lim_{\Delta t \to 0} \prod_{N=1}^{t / \Delta t} e^{-i\hat{H} \Delta t}|\psi_0\rangle,
\end{equation*}
where a second-order symmetric splitting scheme is employed for each time step $\Delta t$
\begin{equation*}
\label{eq:SPO_splitting}
e^{-i\hat{H}\Delta t} = e^{-i\hat{V}_{\text{total}}\Delta t/2} e^{-i\hat{T}\Delta t} e^{-i\hat{V}_{\text{total}}\Delta t/2} + O(\Delta t^3).
\end{equation*}

The numerical procedure for each time iteration $n$ is as follows:
\begin{enumerate}[(a)]
    \item Coordinate space evolution (first half-step).
    \begin{equation}
        \label{eq:step1}
        \psi_{1}(x,n\Delta t) = e^{-i\left[V\left(x,n\Delta t+\frac{\Delta t}{2}\right)+g|\psi(x,n\Delta t)|^{2}\right]\frac{\Delta t}{2}} \psi(x,n\Delta t).   \nonumber
    \end{equation}
    \item Fourier transform to momentum space.
    \begin{equation}
        \label{eq:step2}
        \Phi_{1}(k,n\Delta t) = \frac{1}{\sqrt{2\pi}}\int_{-\infty}^{\infty} \psi_{1}(x,n\Delta t) e^{-ikx}  dx. \nonumber
    \end{equation}
    \item Kinetic evolution in momentum space.
    \begin{equation}
        \label{eq:step3}
        \Phi_{2}(k,n\Delta t) = e^{-i\frac{k^{2}}{2}\Delta t} \Phi_{1}(k,n\Delta t).  \nonumber
    \end{equation}
    \item Inverse Fourier transform to coordinate space.
    \begin{equation}
        \label{eq:step4}
        \psi_{2}(x,n\Delta t) = \frac{1}{\sqrt{2\pi}}\int_{-\infty}^{\infty} \Phi_{2}(k,n\Delta t) e^{ikx} dk.   \nonumber
    \end{equation}
    \item Coordinate space evolution (second half-step).
    \begin{equation}
        \label{eq:step5}
        \psi_{3}(x,n\Delta t) = e^{-i\left[V\left(x,n\Delta t+\frac{\Delta t}{2}\right)+g|\psi(x,n\Delta t)|^{2}\right]\frac{\Delta t}{2}} \psi_{2}(x,n\Delta t).   \nonumber
    \end{equation}
    \item Update $\psi(x, (n+1)\Delta t)$.
    The wave function at the next time step is updated as follows.
    \begin{equation}
        \label{eq:wavefunction_update}
        \psi(x, (n+1)\Delta t) = \psi_{3}(x, n\Delta t).    \nonumber
    \end{equation}
\end{enumerate}
This procedure allows us to obtain the wave function at an arbitrary future time.

\begin{figure}[tb]
    \centering
    \subfloat{\includegraphics[width=8.8cm,height=7.31cm]{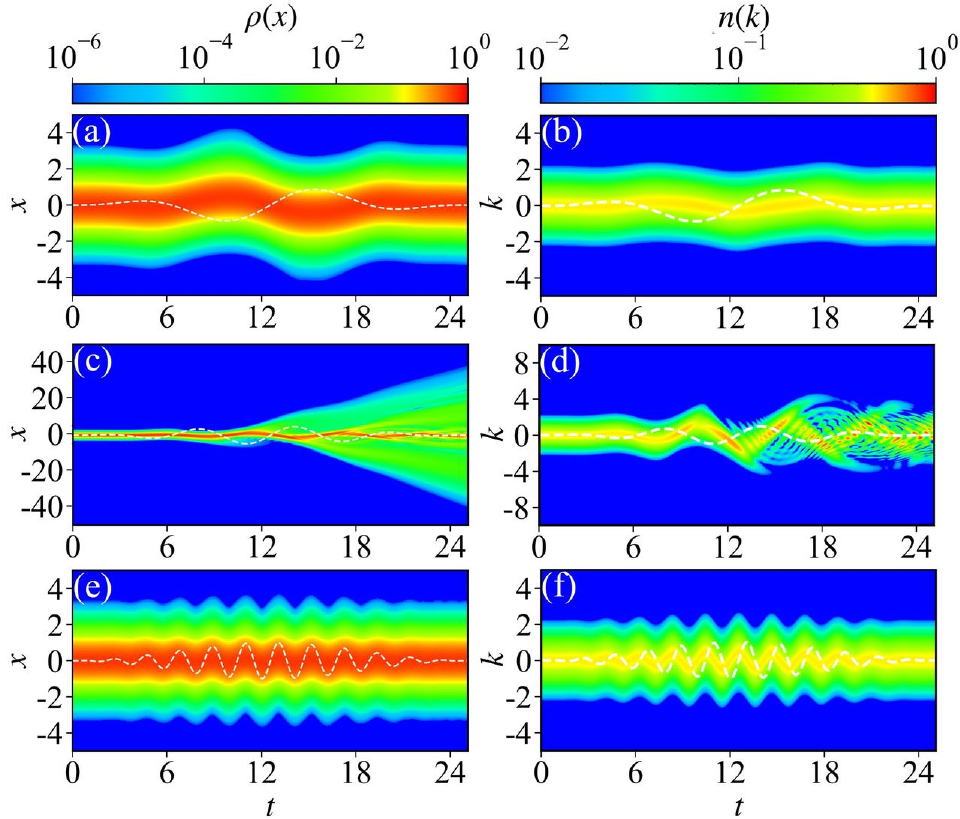}}
    \vspace{0in}
    \caption{Evolution of density distribution in real space and momentum space for different cycle numbers $n_c$. The calculation parameters are set as $F=1$, $g=1$. From top to bottom, the cycle numbers are (a) and (b) $n_c=2$, (c) and (d) $n_c=4$, (e) and (f) $n_c=12$. In this and all subsequent figures, the white dashed line represents the driving field $E(t) = F \sin^2 (\omega t/2 n_c) \sin(\omega t)$.} 
    \label{fig1}
\end{figure}

\begin{figure}[tb]
    \centering
    \subfloat{\includegraphics[width=8.8cm,height=7.31cm]{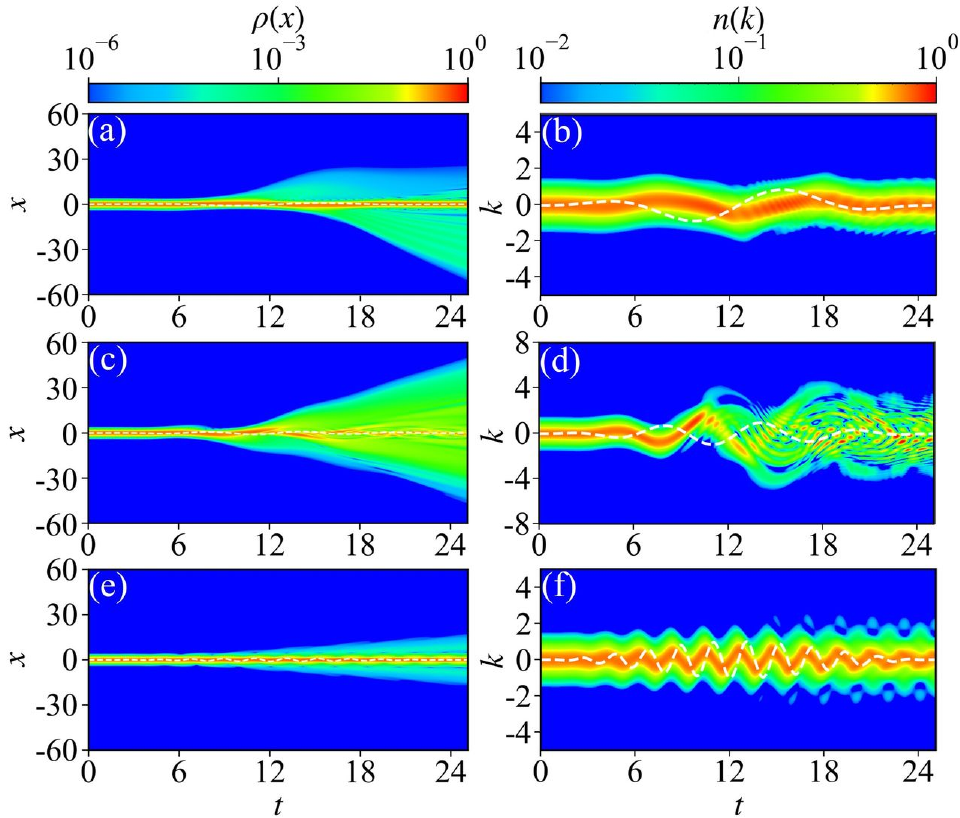}}
    \vspace{0in}
    \caption{Evolution of the density distribution and momentum distribution with $F=1$ and $g=10$. (a) and (b) correspond to $n_c=2$, (c) and (d) $n_c=4$, (e) and (f) $n_c=12$. To enhance the visibility of the driving field curves, the amplitude of the driving field in the density distribution plots has been amplified by a factor of 10. This adjustment applies specifically to the white dashed lines in panels (a), (c), and (e), while the driving field curves in the momentum distribution plots remain unchanged.}
    \label{fig2}
\end{figure}

\begin{figure}[tb]
    \centering
    \subfloat{\includegraphics[width=8.8cm,height=7.31cm]{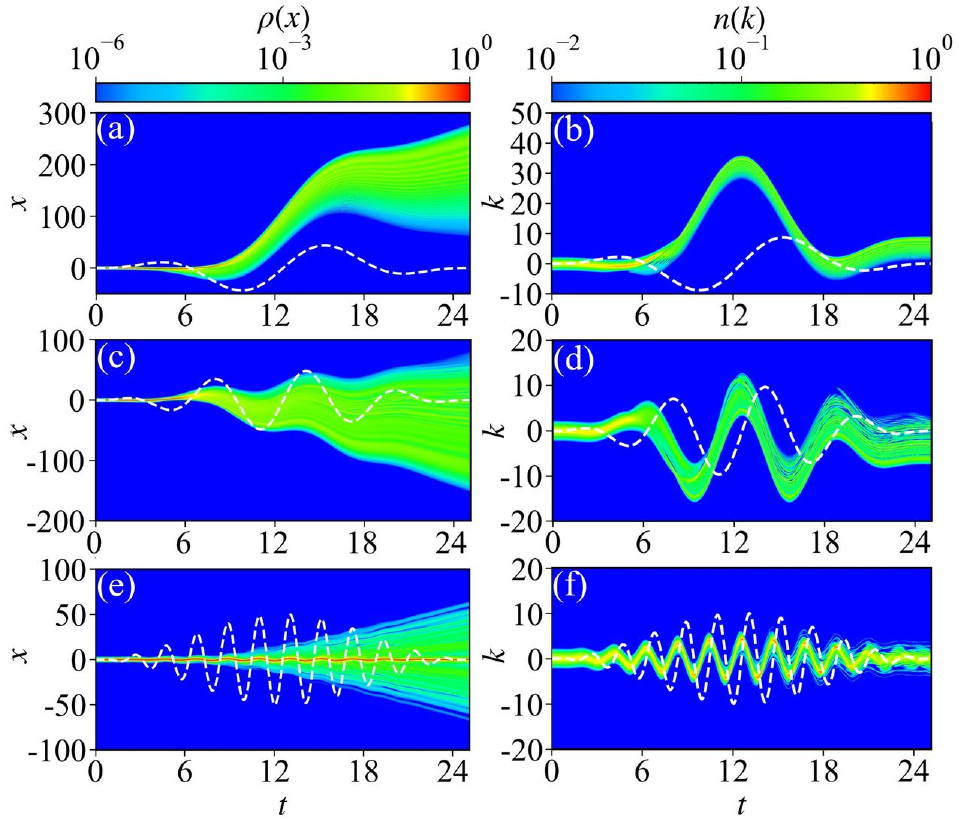}}
    \vspace{0in}
    \caption{Evolution of the density distribution and momentum distribution with $F=10$ and $g=1$. (a) and (b) correspond to $n_c=2$, (c) and (d) $n_c=4$, (e) and (f) $n_c=12$. To enhance the visibility of the driving field curves, the amplitude of the driving field in the density distribution plots has been amplified by a factor of 5. This adjustment applies specifically to the white dashed lines in panels (a), (c), and (e).}
    \label{fig3}
\end{figure}

\section{Evolution of density profile and momentum distribution}

The probability density in real space is defined as
\begin{equation*}
\rho(x, t)=|\psi(x, t)|^{2}.
\end{equation*}
And momentum space density $n(k, t)$ is obtained via Fourier transform
\begin{equation*}
\psi(k, t)=\int\psi(x, t) e^{-i k x} dx,
\end{equation*}
and
\begin{equation*}
n(k, t)=|\psi(k, t)|^{2}.
\end{equation*}

Varying the driving field parameters ($F$, $n_c$) and the atomic interaction strength $g$, we investigate the evolving density distributions. We first plot the evolution of probability density distributions in both the real space and the momentum space of weakly interacting BECs ($g$=1 here) under the pulse driving field with amplitude $F$=1 and different numbers of cycles within the pulse envelope in Fig. \ref{fig1}. It is shown that for specific cycle number ($n_c=4$ here, in Fig. \ref{fig1}(b)) the real space density distribution began to show significant diffusion after the half cycle of the driving field, indicating that some BECs oscillated outside the finite depth potential well, becoming continuum states \cite{Levy2023}. In momentum space, we observed a phenomenon similar to an interference pattern, as shown in Fig. \ref{fig1}(d). With the appearance of non-bound atom the momentum distributions show behaviours similar to the double-slit interference pattern of light. Atoms appear in some regions with larger probability and in other regions with smaller probability. This is markedly different evolution properties from those of a Bose gas in a harmonic potential (which possesses an infinite number of bound-state energy levels), in which case Bose atoms always exhibit a single peak structure in momentum space regardless of how the parameters of the external driving field are varied \cite{Li2025,Hao2023}.
 
Keeping the driving field amplitude $F=1$ and the BECs interaction strength $g=1$, we reduced the number of cycles within the pulse envelope to $n_c=2$. In this case, all atoms always behave as a whole, and the real space density distribution oscillates closely resembled the driving field (the white dashed line in Fig. \ref{fig1}(a)), while the momentum distribution also preserves the single peak structure and exhibits an oscillation similar to the driving field. When the number of cycles within the pulse envelope $n_c$ increases to 12, the evolving characteristics were similar to those of $n_c=2$, and no BECs appeared outside the potential well. This indicates that for the driving field amplitude $F$=1 and the interaction strength $g$=1, if the envelope number is too large or too small, i.e., the driving field oscillates very rapidly or very slowly, BECs will only oscillate in the potential well with the same pattern as the driving field \cite{Li2025,Hao2023}.

In Fig. \ref{fig2} we show the evolving dynamics of the density distribution and the momentum distribution of the BECs with the stronger atomic interaction ($g$=10) exposed to the same pulse drive as the former. In this situation, the BECs  will not only stay in the potential well but also have probabilities of appearing outside it for the strong repulsion between atoms. In addition to the oscillation behaviours as a whole in real space and in momentum space similar to those in the case of weak interaction, more atoms in BECs are excited outside of the potential well and BECs show more remarkable diffusion. In momentum space atoms appear in regions of large momentum.

In Fig. \ref{fig3} we plot the evolving density distribution and the momentum distribution for the driving field with a larger amplitude ($F$=10), which is beyond the potential well. Driven by the pulse field, the atoms initially confined in the well no longer distribute only in a finite space but expand in the regime away from the potential well and diffuse in the whole space. For a small number of cycles ($n_c$=2), the BECs as a whole are pulled out of the trap and gradually diffuse in free space (Fig. \ref{fig3}(a)). With increasing cycle number ($n_c$=4) cold atoms oscillate around the potential well and diffuse simultaneously throughout the space (Fig. \ref{fig3}(c)). As the pulse field oscillates sufficiently rapidly ($n_c$=12), most atoms remain confined in the potential well and only a small number of atoms expand outside the well (Fig. \ref{fig3}(e)). In momentum space atoms behave as a single peak structure and oscillate as a whole body with the variation of driving field in the first half of the pulse drive. In the second half of the pulse the oscillation amplitude begins to decrease and diffusion phenomenon appears.

In short, it has been shown that increasing the amplitude of the driving field or increasing the atomic interaction strength leads to diffusing of BECs, which is the same as the unbinding dynamics \cite{NATURE_COMMUNICATIONS_9_2065(2018)}. But when the pulse drive oscillates rapidly enough, the BECs are bounded in a finite momentum regime and the diffusion phenomena in real space are also suppressed.

\section{Evolution of occupation probability of bound and continuum states}

In order to investigate diffusion phenomena, we expand the above time-dependent wavefunction $\psi(x,t)$ with the eigen states of the Gaussian potential as \cite{ArguelloLuengo2021}
\begin{equation*} 
\psi(x, t)=\sum_{n} C_{n}(t)\phi_{n}(x) e^{-i E_n t}+\int C_{s}(t)\varphi_{s}(x) e^{-i E_s t} dE_s,
\end{equation*}
where $\phi_n(x)$ and $\varphi_s(x)$ are bound state eigenfunction ($E<0$) and continuum state eigenfunction ($E>0$), respectively, which are obtained by diagonalizing the stationary Schrodinger equation (see Appendix A for details). The coefficients $C_n(t)$ can be obtained by calculating the inner product of the time-dependent wave function and the bound states as follows
\begin{align*}
C_{n}(t)&=\int \phi_{n}^*(x) \psi(x, t)dx,
\end{align*}
and $P_{\text{bound}}(t)= \left|C_{n}(t)\right|^{2}$ is the occupation probability of the corresponding eigenstate. The occupation probability of continuum states can also be evaluated with the same procedure and formulated as $P_{\text{continuum}}(t)= \left|C_{s}(t)\right|^{2}$ with $C_{s}(t)=\int\varphi_{s}^*(x)\psi(x, t) dx$.

In the above calculation, the numerical accuracy of the continuum state is crucial and we improve it by requiring the coordinate space to be large enough and verifying the rationality of our continuum states with the following test condition \cite{Geltman1977,CollinsMerts1988}:
\begin{align*}
\sum P_{\text{continuum}}(t) = 1- \sum P_{\text{bound}}(t).
\end{align*}

\begin{figure}[tb]
    \centering
    \subfloat{\includegraphics[width=8.6cm,height=6.45cm]{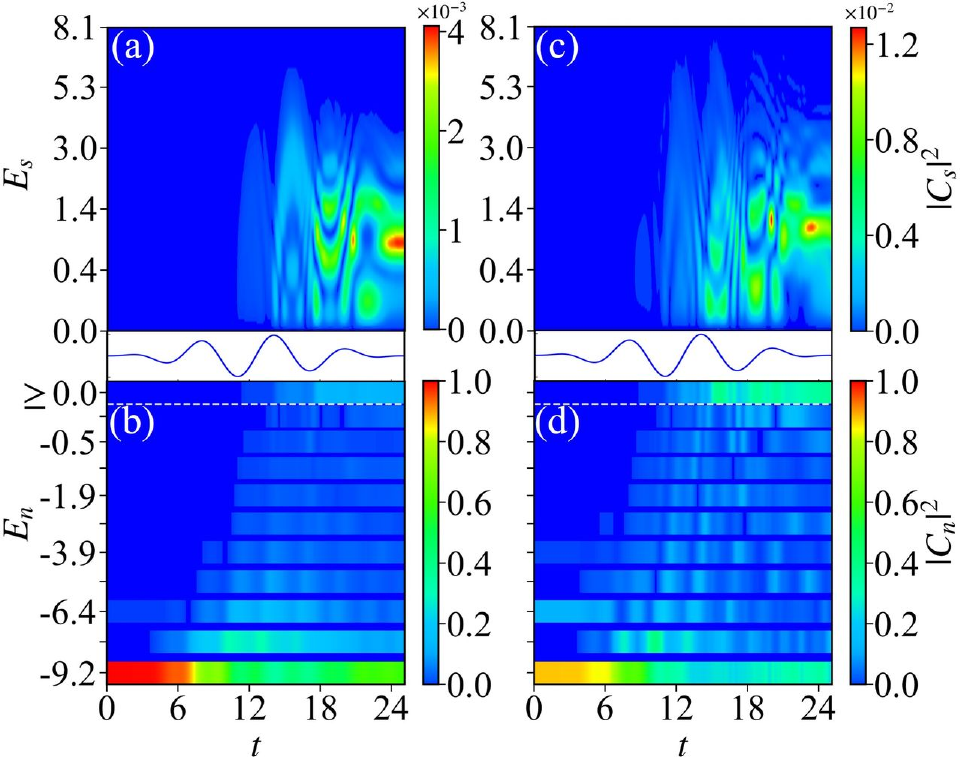}}
    \vspace{0in}
    \caption{Evolution of the occupation distribution for continuum states and bound states under different interaction strengths, with $F=1$ and $n_c=4$. (a) and (b) show the evolution of occupation distribution for continuum states and bound states, respectively, at $g=1$. (c) and (d) show the evolution of occupation distribution for continuum states and bound states, respectively, at $g=10$. The vertical axis represents the energy of different eigenstates. In (b) and (d), there are a total of 10 discrete bound state energy levels. The portion with energy $ \ge 0$ represents the sum of occupation numbers of all continuum states.}
    \label{fig4}
\end{figure}

\begin{figure}[tb]
    \centering
    \subfloat{\includegraphics[width=8.6cm,height=6.45cm]{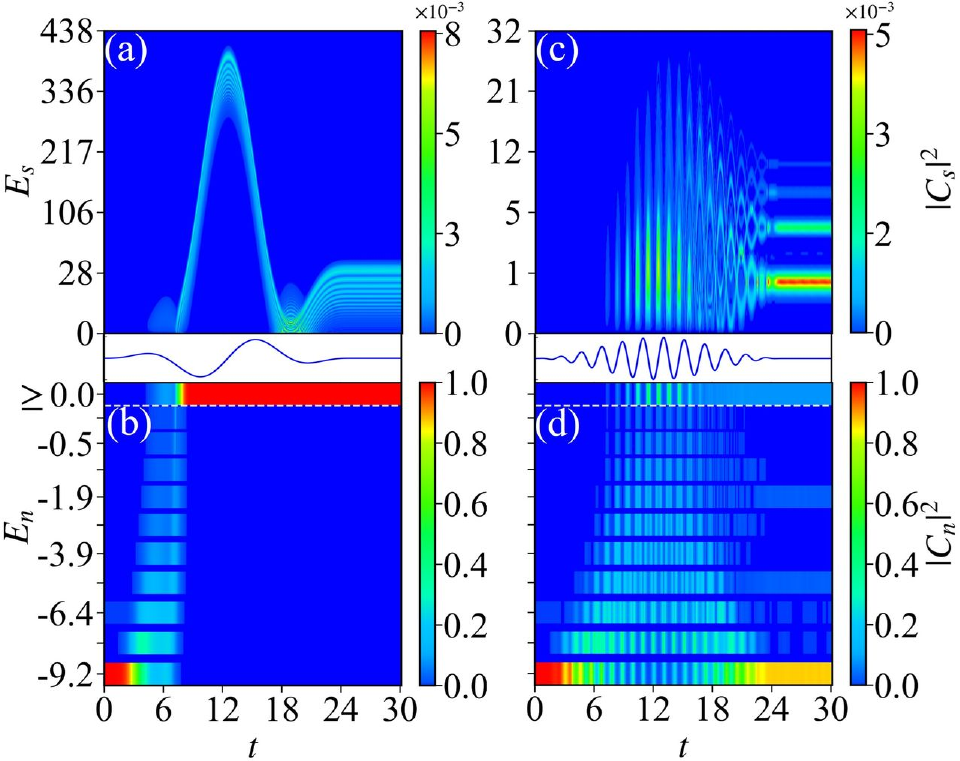}}
    \vspace{0in}
    \caption{Evolution of the occupation distribution for continuum states and bound states under different cycle numbers, with $F=10$ and $g=1$. (a) and (b) show the evolution of occupation distribution for continuum states and bound states, respectively, at $n_c=2$. (c) and (d) show the evolution of occupation distribution for continuum states and bound states, respectively, at $n_c=12$.  The plotting method used in this figure is the same as that in Fig. \ref{fig4}.}
    \label{fig5}
\end{figure}

\begin{figure}[tb]
    \centering
    \subfloat{\includegraphics[width=8.0cm,height=6.3cm]{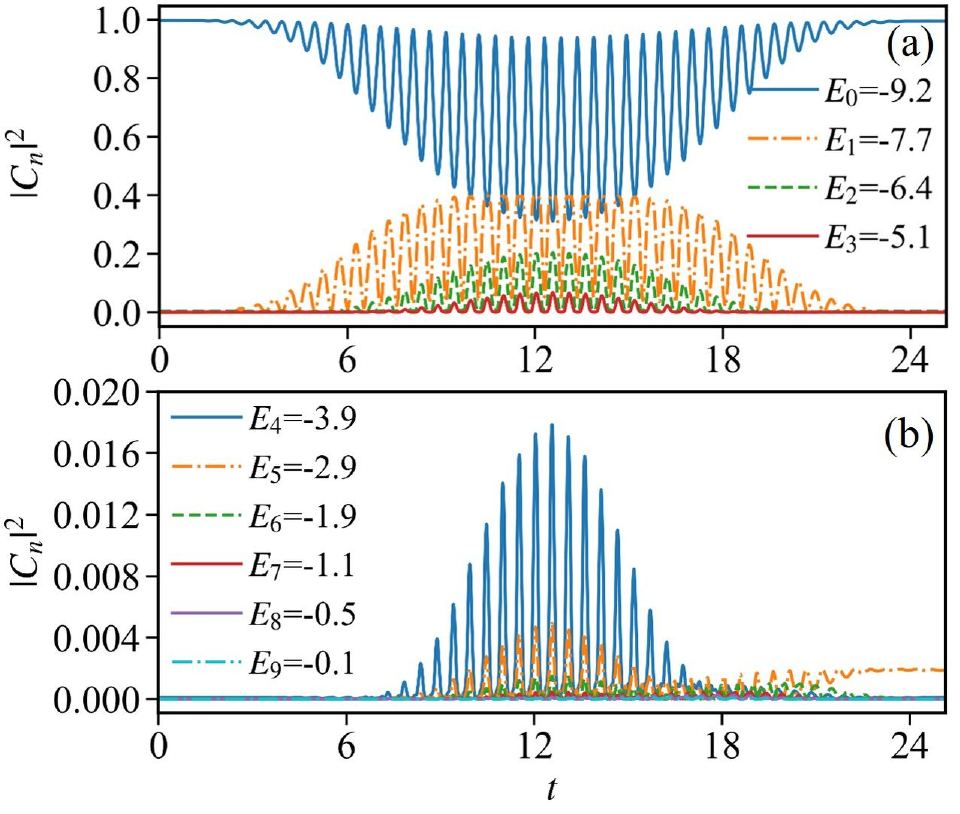}}
    \vspace{0in}
    \caption{The evolution of occupation numbers for bound-state energy levels, with the calculation parameters set as $F=10$, $n_c=24$, and $g=1$. (a) Occupation of the ground state and the 1st to 3rd excited states. (b) Occupation of the 4th to 9th excited states.\\
    }
    \label{fig6} 
\end{figure}

Fig. \ref{fig4} and Fig. \ref{fig5} illustrate the evolution of the occupation probability of the continuum state ($E_s$, (a) and (c)) and the bound state ($E_n$, (b) and (d)) for different BECs and pulse driving fields. In Fig. \ref{fig4} we fix the amplitude of the driving field $F=1$ and the number of cycles within the pulse envelope $n_c$=4, and the atomic interaction strength $g=1$ (Fig. \ref{fig4}(a) and (b)) and 10 (Fig. \ref{fig4}(c) and (d)), respectively. The total occupation of continuum states is shown in the tops of (b) and (d). It is shown that as BECs are driven by a pulse, atoms begin to transit from the ground state to excited states gradually, and more and more atoms populate at high levels until the continuum states ($E>0$ region). During the pulse drive, the occupation of continuum states displays the oscillation behaviour. At the end of the pulse the stable occupation distributions are arrived at. The stronger atomic interaction also induces a larger occupation of continuum states. As shown in Fig. \ref{fig4}, the energy of the continuum shows that there is a concentrated occupation of the final state energy which is approximately 0.4 to 1.4. As the strength of the interaction increases to $g=10$, although the distribution of energy becomes more uniform, the phenomenon of concentrated occupation of the continuum energy still occurs.

Similarly, when the pulse drive is strong and oscillates rapidly ($F=10$ and $n_c=12$, Fig. \ref{fig5} (c)), we also finally observed the phenomenon of energy concentration, although atoms oscillate between bound states and continuum states during the pulse drive. By comparing the occupation of the continuous states at the end of the driving pulse with that at the midpoint of the pulse, we observe that while the occupations of the continuum states at the pulse end are lower than those at the midpoint, the occupation of the continuum state become more concentrated within a specific energy range.

When the driving field is strong ($F$=10) and oscillates slowly, i.e. the number of cycles within the pulse envelope is small ($n_c$=2), the density distribution in the real space shows that all atoms are pulled out of the potential well during the pulse driving process (Fig. \ref{fig3}(a)). The occupation distribution of the continuum and bound states indicates that all atoms have transitioned to the continuum state at $t \approx 8$ and transited between different continuum states after that (Fig. \ref{fig5}(a) and Fig. \ref{fig5}(b)). After the driving field pulse ends, the occupation distribution of the continuum state continues to remain stable. 

As illustrated in the bound-state occupation plot in Fig. \ref{fig6}, the probability that the system occupying continuum states becomes nearly negligible when the pulse drive oscillates with high frequency ($n_c$=24). Instead, the dynamics are dominated by a redistribution of atoms among the bound states, with significant changes in occupation occurring only for the first three levels. In contrast, the occupation probabilities for higher excited states (from the 6th to the 10th level) remain very low, never exceeding 0.01. Notably, after the driving pulse ends, the BEC almost entirely returns to the lowest level.

Another particularly interesting phenomenon is that the frequency of the driving field influences the transition patterns of the particles. In Figs. \ref{fig4} and \ref{fig5}, the atoms consistently prefer to transit to the next-nearest energy levels, indicating that the transition rate between even-even energy levels is greater than that between even-odd levels. This is illustrated in the figures by the fact that the second excited state energy level is the first to exhibit significant particle occupation. This is most evident in Fig. \ref{fig4}(d), where even the fourth excited state is occupied prior to the first excited state. However, when the frequency of the driving field increases further (as shown in Fig. \ref{fig6} for $n_c$=24), the transition pattern reverses: atoms preferentially transition to the nearest neighbour energy levels, and the transition rate from the ground state to excited states decreases as the energy level increases. This is indicated in the figures by the first excited state being the first level to be occupied, and at the same time, higher energy levels exhibit smaller occupation numbers. The transition pattern under these conditions closely resembles that of a harmonic trap. This occurs because, when the driving field frequency is sufficiently high, the expansion of the condensate (or the transition of particles to higher energy states) is severely restricted, confining it largely to the center of the potential well. Under such conditions, the potential well function can be approximated by a low-order expansion ${V_{{\rm{ext}}}}(x) \approx {V_0}\left( {{{{x^2}} \mathord{\left/{\vphantom {{{x^2}} {2r_0^2}}} \right. \kern-\nulldelimiterspace} {2r_0^2}} - 1} \right)$, effectively resembling a harmonic trap.

This occupation behavior strongly correlates with the real- and momentum-space density evolution discussed before. When $n_c$ is excessively large while the total driving time $T$ is constant, the oscillation frequency of the driving field increases significantly. As a result, the BEC undergoes more cycles within the same duration, which effectively suppresses the quantum diffusion phenomenon by inhibiting the transition of atoms into continuum states. Thus, the system remains predominantly confined within the bound levels, consistent with the observed inhibition of spatial expansion in the density profiles.

\section{HHG yield}

Following the Larmor formula, the HHG yield is proportional to the square of dipole acceleration magnitude. The photon energy spectrum is given by the Fourier transform \cite{Tong2023}
\begin{equation*}
d_{a}(\Omega)=\int d_a(t) e^{-i\Omega t} dt,
\end{equation*}
Dipole acceleration is computed via Ehrenfest's theorem
\begin{equation*}
d_{a}(t)=\langle\psi(t)|(-\nabla V_{\text{ext}}(x))|\psi(t)\rangle,
\end{equation*}
with $\nabla V_{\text{ext}}$ denoting the gradient of the confining potential. Taking into account the time interval of dipole acceleration $(0, T)$, where $T$ is the period of the driving field, the formula to calculate the dipole acceleration spectrum becomes \cite{Krause1992,Xia2022}
\begin{equation*}
d_{a}(\Omega)=\frac{1}{T}\int_0^T d_a(t) e^{-i\Omega t} dt. 
\end{equation*}

\begin{figure}[tb]
    \centering
    \subfloat{\includegraphics[width=8.5cm,height=4.72cm]{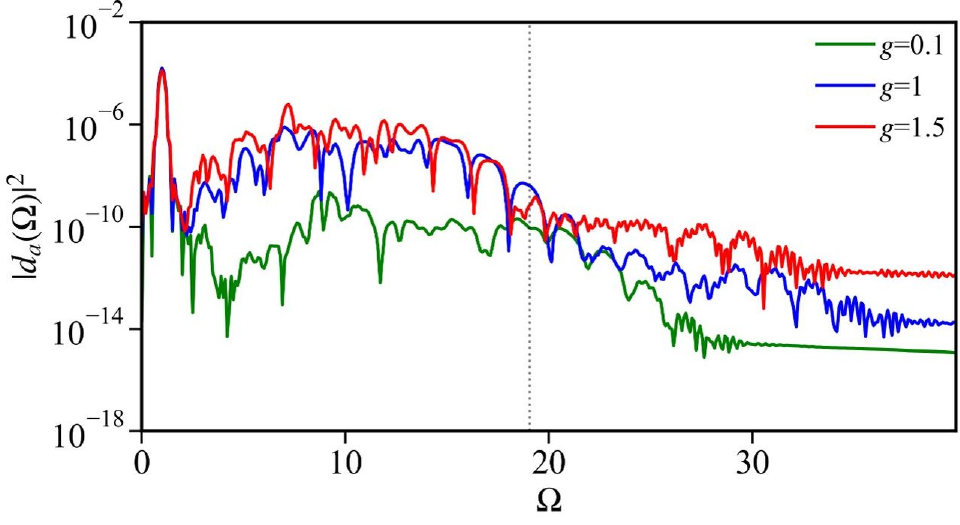}}
    \vspace{0in}
    \caption{The HHG yield for different repulsive interaction strengths $g$ with typical cutoff frequency characteristics. The parameters of driving field are as follows: $F=0.05$ and $n_c$=6. The gray dashed line corresponds to the cutoff frequency $\Omega_c=I_p+3.17U_p$. Here $I_p$ is the ionization potential of Gaussian potential and $U_p = F^2/(4\omega^2)$ is the ponderomotive energy \cite{PRX_QUANTUM_5_010328(2024)}.}
    \label{fig7}
\end{figure}

The HHG yields for different atomic interactions are displayed in Fig. \ref{fig7}. It is shown that the yield spectrum displays the main feature of the harmonic spectrum: a rapid decline in the low-order harmonics followed by a plateau where the harmonic yield is almost constant and then an abrupt decline. The HHG cutoff frequency $\Omega_c=I_p+3.17U_p=19$ is plotted in dashed lines. It is shown that the atomic interaction results in the shift of the cutoff frequency to a smaller or larger location. Here, the cutoff frequency is defined as the position at which the HHG yield declines abruptly following the plateau. It is also interesting to note that in the HHG yield spectrum, after the constant platform followed by an abrupt decrease, another platform follows closely behind. Atomic repulsion induces the higher constant plateau, i.e., the higher HHG yield, because of the higher energy of BECs. As the interaction changes from 0.1 to 1, the yield enhances by several orders of magnitude.

\section{Conclusion}

In conclusion, the response dynamics of BECs to strong oscillatory pulse drive are investigated by numerically solving GPE. The initial states are prepared with the imaginary time evolution method by the Crank-Nicholson discretization scheme, and the time-dependent wavefunction is obtained with SPO method. We obtained the evolving dynamics of the density distribution, momentum distribution, occupation distribution in energy space, and HHG yield. For the weakly interacting BECs, all atoms oscillate as a whole body together with the pulse drive and always display a typical single peak structure when the pulse field is weak and change very slowly or rapidly. As the oscillation frequency is suitable, some atoms are excited out of the potential well and diffuse in the whole space. The stronger the atomic interaction, the more obvious the diffusion. When the amplitude of pulse drive goes beyond the potential well, more atoms are pulled out of the potential and diffuse in the whole real space but still distribute in the finite momentum region. This phenomenon can be suppressed as the cycle number of the pulse drive is high enough. The diffusion in the real space and the distribution in the finite momentum region can be understood by the occupation in the energy space. When the pulse amplitude and the atomic interaction are strong enough, atoms are excited into the continuum states but still concentrate in the specific energy region. When the pulse drive oscillates rapidly enough, that atoms have an ignorable probability to occupy the continuum state and will always be confined in the potential well. It is also shown that the atomic yield can enhance the HHG yield by several orders of magnitude as the interaction increases tenfold.

\begin{acknowledgements}
R. J. is support by the Young Scientists Fund of the National Natural Science Foundation of China (Grant No. 12404296) and Original Exploration Program of the National Natural Science Foundation of China (Grant No. 12450404).
\end{acknowledgements}

\noindent{\bf Data availability} \\
The data that support the findings of this article are openly available.

\appendix

\section{Calculation of Energy Eigenstates}
\label{subsec:eigenstate_calculation}

To analyze the occupation probabilities of bound and continuum states during the dynamics, we numerically solve the stationary Schrodinger equation for the Gaussian potential:
\begin{equation}
\label{eq:stationary_schrodinger}
-\frac{1}{2}\frac{d^{2}}{dx^{2}}\phi(x) - V_{0}\exp\left(-\frac{x^{2}}{2 r_{0}^{2}}\right)\phi(x) = E\phi(x).
\end{equation}

Discretizing on a spatial grid with spacing $\Delta x$ using finite differences yields the Hamiltonian matrix:
\begin{equation}
\label{eq:hamiltonian_matrix}
\hat{H} = \begin{bmatrix}
2A + V(x_0) & -A & 0 & \cdots \\
-A & 2A + V(x_1) & -A & \cdots \\
0 & -A & 2A + V(x_2) & \\
\vdots & \vdots & & \ddots
\end{bmatrix},
\end{equation}
where $A = 1/[2(\Delta x)^2]$ and $V(x_i) = -V_{0}\exp\left(-x_i^{2}/(2 r_{0}^{2})\right)$. We take 8001 discrete points in space, a spatial step $\Delta x$ length of 0.05. Full diagonalization of $\hat{H}$ provides the complete set of energy eigenstates (bound and continuum), which are used to compute projections of the time-dependent wave function.

\bibliography{apssamp}

\end{document}